\documentclass[reprint,aps,superscriptaddress,amsmath,amssymb,longbibliography]{revtex4-1}
\usepackage{graphicx}% Include figure files
\usepackage{dcolumn}% Align table columns on decimal point
\usepackage{bm}% bold math                                     
\usepackage{color}
\usepackage{xspace}                     
\usepackage{ulem}
\usepackage[colorlinks=true,urlcolor=blue,citecolor=blue,linkcolor=blue,bookmarks=false,
pdfstartview={FitH}]{hyperref}

\begin{document}

\title{On the origin of critical nematic fluctuations in pnictide
superconductors}   
\author{S.-F. Wu}\email{sfwu@iphy.ac.cn}
\affiliation{Department of Physics and Astronomy, Rutgers University,
Piscataway, NJ 08854, USA}
\affiliation{Beijing National Laboratory for Condensed Matter
Physics, and Institute of Physics, Chinese Academy of Sciences,
Beijing 100190, China}

\affiliation{School of Physical Sciences, University of Chinese
Academy of Sciences, Beijing 100190, China}          
\author{ W.-L. Zhang}
\affiliation{Department of Physics and Astronomy, Rutgers University,
Piscataway, NJ 08854, USA}
\author{L. Li}
\affiliation{Materials Science \& Technology Division, Oak Ridge
National Laboratory, Oak Ridge, TN 37831}
\author{H. B. Cao}
\affiliation{Neutron Scattering Division, Oak Ridge National
Laboratory, Oak Ridge, TN 37831}       
\author{ H.-H. Kung}
\affiliation{Department of Physics and Astronomy, Rutgers University,
Piscataway, NJ 08854, USA}             
\author{A. S. Sefat}    
\affiliation{Materials Science \& Technology Division, Oak Ridge
National Laboratory, Oak Ridge, TN 37831}
\author{H. Ding}
\affiliation{Beijing National Laboratory for Condensed Matter
Physics, and Institute of Physics, Chinese Academy of Sciences,
Beijing 100190, China}
\affiliation{School of Physical Sciences, University of Chinese
Academy of Sciences, Beijing 100190, China}
\affiliation{Collaborative Innovation Center of Quantum Matter,
Beijing, China}
\author{P. Richard}\email{pierre.richard.qc@gmail.com}
\affiliation{Beijing National Laboratory for Condensed Matter
Physics, and Institute of Physics, Chinese Academy of Sciences,
Beijing 100190, China}
\affiliation{School of Physical Sciences, University of Chinese
Academy of Sciences, Beijing 100190, China}
\affiliation{Collaborative Innovation Center of Quantum Matter,
Beijing, China}
\author{G. Blumberg}\email{girsh@physics.rutgers.edu}
\affiliation{Department of Physics and Astronomy, Rutgers University,
Piscataway, NJ 08854, USA}
\affiliation{National Institute of Chemical Physics and Biophysics,
12618 Tallinn, Estonia}
\date{\today}

\begin{abstract}
We employ polarization-resolved Raman spectroscopy to study critical nematic 
fluctuations in Ba(Fe$_{1-x}$Au$_x$)$_2$As$_2$ 
superconductors above and across well separated tetragonal to orthorhombic phase
transition at temperature $T_S(x)$  and the N\'{e}el
transition at $T_N(x)$.
The static Raman susceptibility in $XY$ symmetry
 channel increases upon cooling from room temperature
following the Curie-Weiss law, with Weiss temperature 
$T_{\theta}(x)$ several tens of degrees lower than $T_S(x)$. 
Data reveals a hidden nematic quantum critical point at $x_{c} = 
0.031$ when the system becomes superconducting,  
indicating a direct connection between quantum critical nematic 
fluctuations and unconventional superconductivity. 
We attribute the origin of the nematicity to charge quadrupole
fluctuations due to electron transfer between the nearly degenerate
$d_{xz}/d_{yz}$ orbitals.

\end{abstract} 
        

\maketitle

It is widely believed that the interactions leading to
high-temperature superconductivity are already present in the 
parent compounds. 
The parent compounds of the Fe-based
superconductors usually show a tetragonal to orthorhombic structural 
transition at  
temperature $T_S$ that is accompanied by transition into 
collinear antiferromagnetic phase at temperature $T_N$, typically 
only slightly lower than $T_S$. 

Recently, much attention was devoted to studies of non-symmetric dynamical 
fluctuations above $T_S$ which break local four-fold symmetry, usually 
referred as nematic fluctuations~\cite{fradkin2010review}. 
Below $T_S$, significant anisotropy was 
found for properties measured along the two planar
orthogonal Fe-Fe directions, notably in electrical
resistivity~\cite{Chu824science2010}, optical
conductivity~\cite{Dusza2011EPL}, thermopower~\cite{Jiang2013PRL110}, 
and local density-of-states (DOS)~\cite{Rosenthal2014NatPhy}. 
It has been established both by the static probes such as shear modulus
$C_{66}$~\cite{GotoJPSJ,Yoshizawa2012JPSJ,Yoshizawa2012ModPhyB,Kurihara2017JPSJ}, 
Young's modulus $Y_{110}$~\cite{Bohmer2014PRL,BOHMER201690}, 
the elastoresistance coefficient 
$m_{66}$~\cite{Kuo958science2016,Chu710science2012}, and by the dynamic 
probe: polarization resolved Raman 
scattering~\cite{Gallais2013PRL,Massat_PNAS113,GALLAIS2016113,Kretzschmar_Nature12,Bohm2017physica,Thorsmolle2016PRB,ZhangWL2014arxiv,Kaneko2017PRB,Zhang2017arxiv}, 
that the underlying nematic fluctuations have a
distinct $XY$ quadrupole symmetry and that they extend to temperatures far 
above $T_S$. 
However, the origin of the nematic fluctuations remains under
debate. 
            
Among interpretations, it has been proposed that the 
fluctuations could originate from charge transfer between
degenerate $d_{xz}/d_{yz}$
orbitals~\cite{Onari2012PRL,Kontani2011PhysRevB,Kontani2014PhysRevLett,Yamase2013PhysRevB,Kruger2009PhysRevB,Lv2011PhysRevB,BASCONES201636,GALLAIS2016113,Khodas2015PhysRevB},
or from magnetic
interactions~\cite{Fernandes2014NatPhy,Fernandes2010PhysRevLett,Fernandes2012PRB,Chubukov2016PhysRevX,INOSOV201660,Karahasanovic2015PhysRevB,Hinojosa2016PhysRevB,Classen2017PhysRevLett}.
We noticed that for Ba(Fe$_{1-x}$Au$_x$)$_2$As$_2$ superconductors the $T_S$ and 
$T_N$ transition temperatures are well separated, thus the system provides a
platform to study separately the charge and spin contributions to the 
nematic fluctuations.

\begin{figure}[!t]
\begin{center}
\includegraphics[width=0.75\columnwidth]{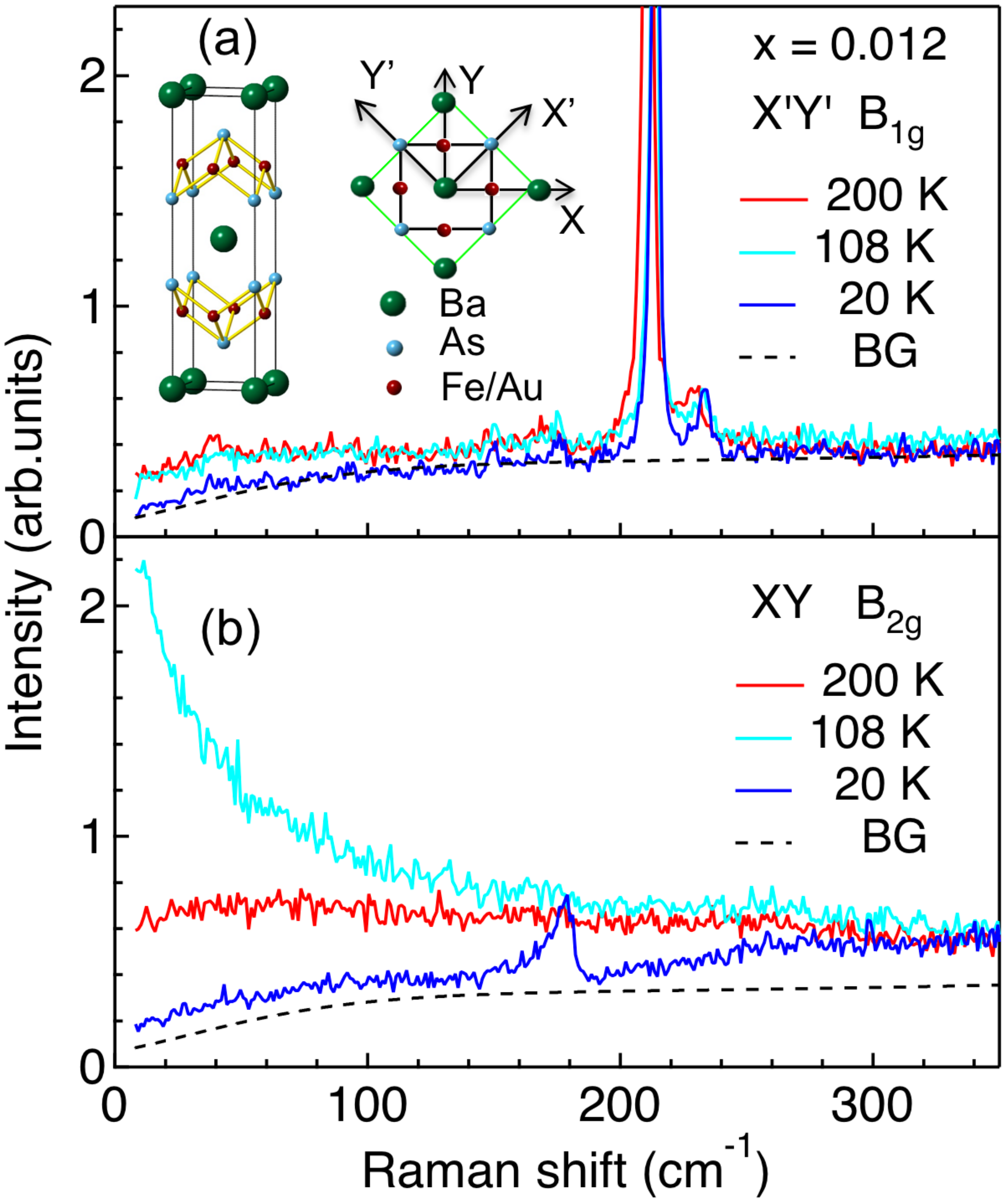}
\end{center}
\caption{\label{Fig1_B1gB2g}
Crystal structure of Ba(Fe$_{1-x}$Au$_x$)$_2$As$_2$ and the definition of  
$X$, $Y$, $X'$ and $Y'$ directions are shown in the top panel. 
The green and black lines 
represent  4-Fe and 2-Fe unit cells, respectively. 
(a) and (b) Secondary emission for $X'Y'$ and $XY$ polarizations, 
correspondingly. 
The dashed lines show background at 20~K determined by intensity in $X'Y'$ 
polarization. 
}     
\end{figure}
\begin{figure*}[!t]
\begin{center}
\includegraphics[width=2\columnwidth]{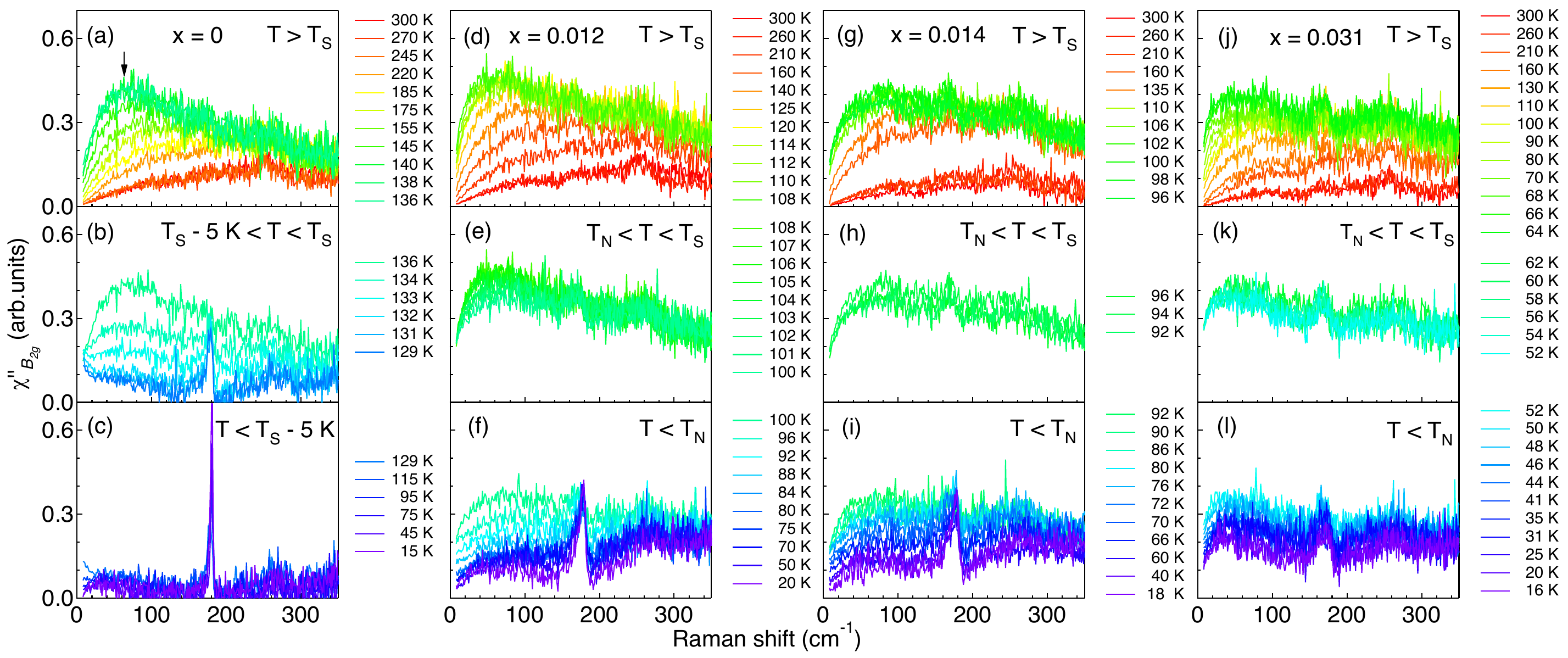}
\end{center}
\caption{\label{Fig2_B2g}Temperature dependence of the Raman response
$\chi''_{B_{2g}}(\omega,T,x)$ for Ba(Fe$_{1-x}$Au$_x$)$_2$As$_2$.
(a)-(c) $x=0$, (d)-(f) $x=0.012$, (g)-(i) $x=0.014$ and (j)-(l)
$x=0.031$. The arrow in (a) indicates the QEP.}
\end{figure*} 
In this Letter, we use the static Raman susceptibility derived from 
the dynamical response acquired  
in the $XY$-symmetry quadrupole channel to study the
evolution of the nematic fluctuations above and across the structural and
magnetic phase transitions as function of Au doping into 
Ba(Fe$_{1-x}$Au$_x$)$_2$As$_2$ crystals. 
Above $T_{S}$, the static Raman susceptibility follows the 
Curie-Weiss law with Weiss temperature $T_{\theta}(x)$ about 40-60~K 
lower than $T_S$. 
The growth of susceptibility stops below $T_{S}$, 
when degeneracy of the $d_{xz}$ and $d_{yz}$ orbitals is lifted, 
emphasizing a relation between nematicity and  
$d_{xz}/d_{yz}$ quadrupole charge/orbital fluctuations. 
Furthermore, we demonstrate that the charge quadrupole moment is 
monotonically increasing with Au doping while the ordered magnetic 
moment is decreasing with the doping, indicating that the charge 
quadrupole fluctuations and stripe magnetic orders are competing. 
Moreover, the data reveal a hidden quantum critical point at 
critical doping $x_{c}$ defined by $T_{\theta}(x_{c})=0$, which 
appears to be at the very heart of the superconducting dome.  
Below $T_N$, the susceptibility decreases rapidly upon cooling  as the magnetic 
order parameter develops and as a spin-density-wave (SDW) gap is 
depleting the DOS of the occupied $d_{xz}/d_{yz}$ orbitals. 
\begin{table}[!b]
\caption{\label{TS} Summary of the structural and magnetic phase
transition temperatures determined by neutron scattering [Figs.
\ref{Fig3_Chi0}(f)-\ref{Fig3_Chi0}(h)], Weiss temperature
$T_\theta$(K), ordered magnetic moment and charge
quadrupole moment $Q(x)$ for Ba(Fe$_{1-x}$Au$_x$)$_2$As$_2$ samples.}
\begin{ruledtabular}
\begin{tabular}{cccccc}
Sample&$T_{S} (K)$&$T_{N}$ (K)&$T_\theta$
(K)&M($\mu_B)$&$Q(x)$(arb.units)\\
\hline
x=0&135&135&96$\pm$4&0.87~\cite{Dai2015RevModPhys}&5.7$\pm$0.2\\
x=0.012&108&100&64$\pm$4&0.50$\pm0.02$~\cite{Au122_phonon_paper}&6.7$\pm$0.3\\
x=0.014&96&92&46$\pm$8&0.42$\pm0.04$~\cite{Au122_phonon_paper}&6.7$\pm$0.5\\
x=0.031&63&54&4$\pm$5&0.36$\pm0.02$~\cite{Au122_phonon_paper}&7.5$\pm$0.3\\
\end{tabular}
\end{ruledtabular}
\begin{raggedright}
\end{raggedright}
\end{table}

\begin{figure*}[t] 
\begin{center}
\includegraphics[width=2\columnwidth]{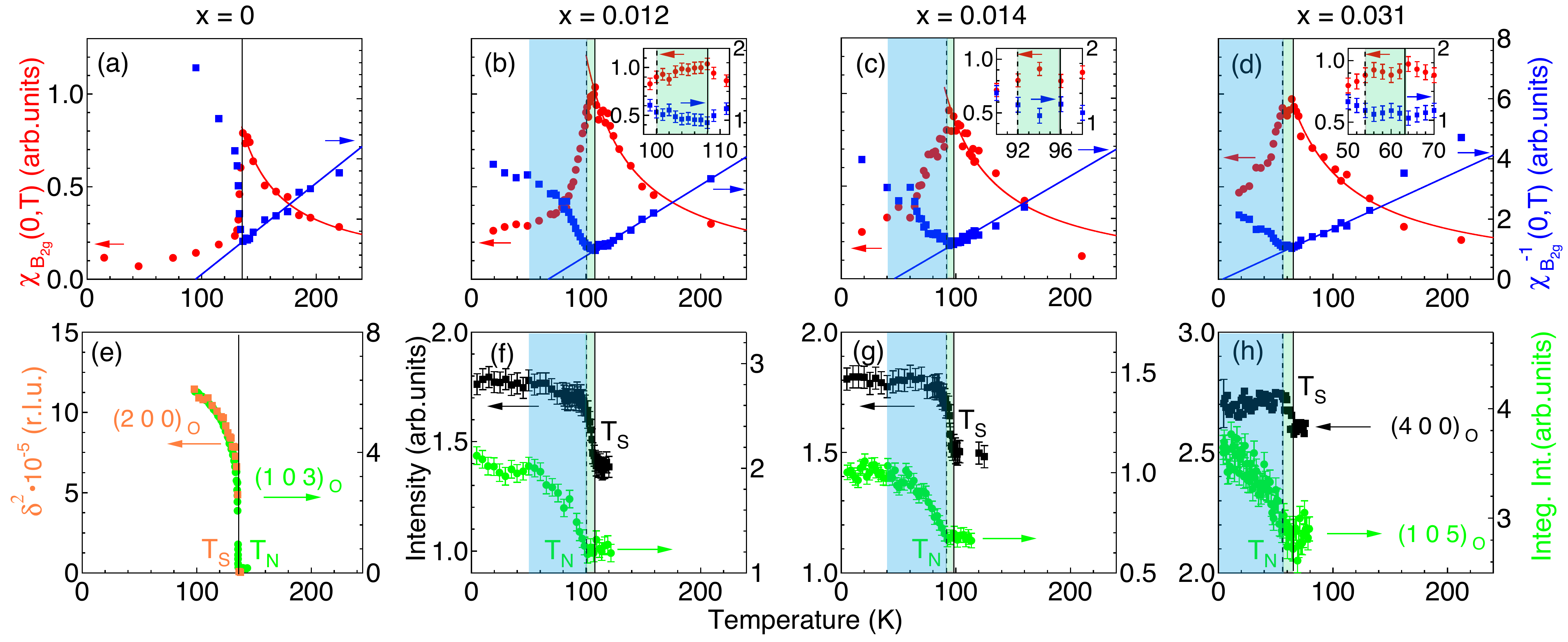}
\end{center}
\caption{\label{Fig3_Chi0}(a)-(d) 
Temperature dependence of the
static Raman susceptibility $\chi_{B_{2g}}(0,T,x)$ (red solid
circles) and its inverse $\chi_{B_{2g}}^{-1}(0,T,x)$ (blue solid
square) at different doping concentrations $x$. 
The red and blue curves are fits to the
Curie-Weiss law (a linear function for inverse susceptibility). The
solid lines mark $T_{S}(x)$, whereas the dashed lines mark
$T_{N}(x)$. 
The insets in (b)-(d) are zoom-in around the transition
temperatures. 
(e) Temperature dependence of $\delta^2$ (orange solid
squares) deduced from the splitting of the (2$\pm \delta$ 0 0)
nuclear Bragg peak of BaFe$_{2}$As$_2$, from
Ref.\,\cite{Wilson2009PRB}, with $\delta/2=(a+b)/(a-b)$ representing
the orthorhombicity of a crystal with in-plane lattice parameters $a$
and $b$. The green solid circles represent the temperature evolution
of the integrated intensity of the (1 0 3) magnetic Bragg peak, from
Ref.\,\cite{Wilson2009PRB}. (f)-(h) Temperature evolution of the (2 0
0) nuclear Bragg peak intensity (black solid squares) and of the (1 0
5) magnetic Bragg peak intensity (green solid circles) for
Ba(Fe$_{1-x}$Au$_x$)$_2$As$_2$ in the orthorhombic
phase~\cite{Au122_phonon_paper}.
}    
\end{figure*}

Single crystals of Ba(Fe$_{1-x}$Au$_x$)$_2$As$_2$ ($x=0$, 0.012,
0.014, 0.031) were grown out of self-flux using a high-temperature
solution growth technique described in
Refs.~\cite{Sefat2013bulk,Li2015PRB}, and the chemical compositions
were determined by energy dispersive spectroscopy (EDS) analysis. 
The room-temperature crystal structure, illustrated in the inset of 
Figs.~\ref{Fig1_B1gB2g}(a), belongs to space group
$I4/mmm$ (point group $D_{4h}$). 
The $T_S$ and $T_N$ of the Au-doped
samples were determined, respectively, by the temperature evolution of
the neutron nuclear and the magnetic Bragg peak intensities shown in
Figs.~\ref{Fig3_Chi0}(f)-\ref{Fig3_Chi0}(h). 
For the pristine compound,
$T_S \simeq T_N=135$\,K, as was determined by bulk
property measurements~\cite{Li2015PRB}, in agreement with neutron
diffraction measurements~\cite{Wilson2009PRB}. The $T_S$ and $T_N$
values for the compositions studied are summarized in Table~\ref{TS}.

The crystals used for Raman scattering were cleaved and positioned in
a continuous helium flow optical cryostat. The measurements were 
performed in a quasi-back scattering geometry along the 
crystallographic $c$-axis using the Kr$^+$ laser line at 647.1\,nm
(1.92\,eV). The excitation laser beam was focused into a
$50\times100$ $\mu$m$^2$ spot on the $ab$-surface, with the incident
power around 10\,mW. The scattered light was collected and analyzed
by a triple-stage Raman spectrometer designed for high-stray light
rejection and throughput, and then recorded using a liquid
nitrogen-cooled charge-coupled detector. Raman scattering intensity
data were corrected for the spectral responses of the spectrometer
and detector. 
The laser heating in the Raman experiments is determined by imaging
the appearance of stripes due to twin domain formation at the
structural phase transition temperature $T_S$
\cite{Kretzschmar_Nature12}. When stripes appear under laser
illumination, the spot temperature is just slightly below $T_S$, thus
$T_S=kP+T_{cryo}$, where $T_{cryo}$ is the temperature of cold helium
gas in the cryostat, $P$ is the laser power and $k$ is the heating
coefficient. By recording $T_{cryo}$ when the stripes appear at
different laser powers, we can deduce the heating coefficient using a
linear fit: $k =1 \pm0.1$ K/mW. 
                                            
We define $X$ and $Y$ directions along the 2-Fe unit cell
basis vectors (at 45$^{\circ}$ from the Fe-Fe direction) in
the tetragonal phase, whereas $X'$ and $Y'$ are along the Fe-Fe
directions, as shown in the inset of Fig.~\ref{Fig1_B1gB2g}(a). 
According to Raman selection rules,  
the $XX$, $XY$, $X'X'$,
$X'Y'$ scattering geometries probe $A_{1g} + B_{1g}$, $A_{2g}+
B_{2g}$, $A_{1g}+B_{2g}$ and $A_{2g}+B_{1g}$ symmetry excitations of 
the $D_{4h}$ point group respectively.
The data in the $X'Y'$($B_{1g}$) symmetry channel barely changes with 
temperature [Fig.~\ref{Fig1_B1gB2g}(a)].  
In contrast, the spectrum in the $XY$($B_{2g}$) symmetry channel show strong 
temperature dependence [Fig.~\ref{Fig1_B1gB2g}(b)]. 
Therefore, we subtract the background signal recorded in
the $X'Y'$ geometry from the data measured in the
$XY$ geometry to obtain $\chi''_{B_{2g}}(\omega,T)$ response 
function~\cite{Gallais2013PRL}.

In Figs.~\ref{Fig2_B2g}(a)-\ref{Fig2_B2g}(c), we show the Raman
response for BaFe$_2$As$_2$ in the $B_{2g}$ channel at 
temperatures between 300\,K and 15\,K. 
The most remarkable feature appearing upon cooling is a quasi-elastic peak
(QEP) that reaches its maximum intensity around $T_{S}/T_{N}$  [Fig.
\ref{Fig2_B2g}(a)]. The QEP is sharply suppressed within 5\,K below
$T_S/T_N$ [Fig. \ref{Fig2_B2g}(b)] and it vanishes at lower
temperatures [Fig. \ref{Fig2_B2g}(c)].
Above $T_{S}(x)$, the Raman response for the Au-doped samples ($x=0.012$, 0.014, 0.031)
shows similar behavior as for the pristine compound 
[Figs.~\ref{Fig2_B2g}(d), \ref{Fig2_B2g}(g) and \ref{Fig2_B2g}(j)].
Unlike for the pristine compound, $T_{S}(x)\neq T_{N}(x)$ in the doped
samples. Interestingly, the passage across $T_{S}(x)$ does not affect
the Raman response significantly, as shown in Figs.
\ref{Fig2_B2g}(e), \ref{Fig2_B2g}(h) and \ref{Fig2_B2g}(k). The
situation is quite different when temperature decreases below
$T_{N}(x)$ [Figs.~\ref{Fig2_B2g}(f), \ref{Fig2_B2g}(i) and
\ref{Fig2_B2g}(l)]. As with the pristine compound, the QEP is
suppressed quickly upon cooling below $T_{N}(x)$.

For better understanding of QEP evolution in the
$B_{2g}$ channel, we compute the static nematic susceptibility
$\chi_{B_{2g}}(0,T,x)$ from the experimental
data using the Kramers-Kronig
transformation~\cite{Gallais2013PRL,Thorsmolle2016PRB,ZhangWL2014arxiv}: 
\begin{equation}
\chi_{B_{2g}}(0,T,x) \approx
\frac{2}{\pi}\int_0^{350~cm^{-1}}\frac{\chi''_{B_{2g}}(\omega,
T,x)}{\omega}d\omega\,.
\end{equation}
The integrant underlines importance of Raman response in 
low-frequency limit.  
We use linear extrapolation for response below 8\,cm$^{-1}$ 
instrumental cut-off.

\begin{figure}[!t]
\begin{center}
\includegraphics[width=0.9\columnwidth]{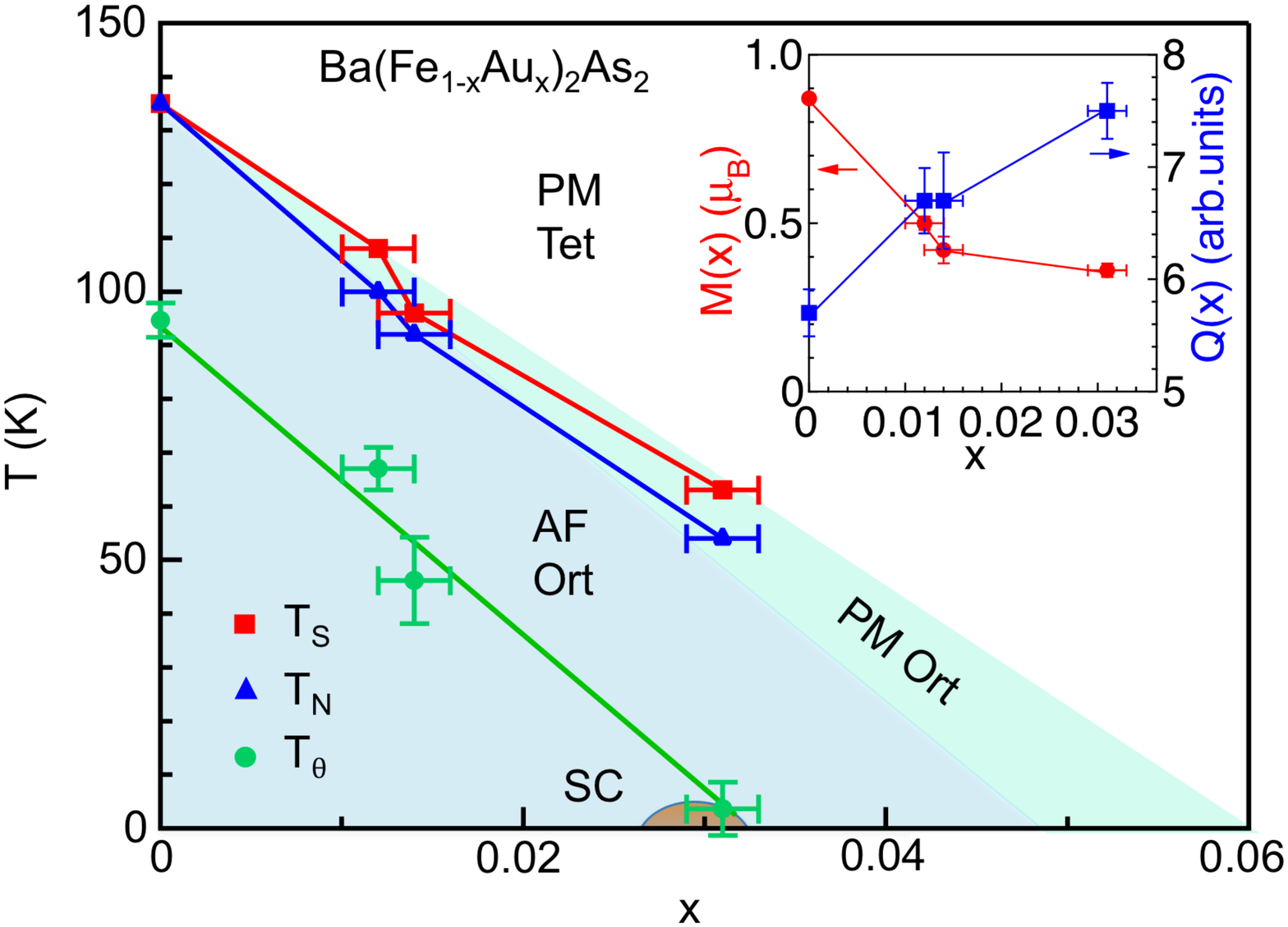}
\end{center}
\caption{\label{Fig4_summary}
Phase diagram of Ba(Fe$_{1-x}$Au$_x$)$_2$As$_2$. 
Red solid squares and blue solid
triangles represent $T_{S}(x)$ and $T_{N}(x)$, respectively. 
The green solid circles denote interpolated Weiss temperature $T_\theta(x)$. 
The inset shows the doping dependence of the ordered
magnetic moment $M(x)$ and charge quadrupole moment $Q(x)$.}      
\end{figure}

In Figs. \ref{Fig3_Chi0}(a)-\ref{Fig3_Chi0}(l), we show 
temperature dependence of $\chi_{B_{2g}}(0,T,x)$. 
Above $T_S(x)$, the static Raman
susceptibility is well described by the Curie-Weiss
law 
\begin{equation}
\label{Curie}
\chi_{B_{2g}}(0,T,x) \propto 
Q^{2}(x)/(T-T_\theta(x)), 
\end{equation}
where $T_\theta(x)$ is the Weiss
temperature and square of the charge quadrupole moment $Q^{2}(x)$ is 
proportional to Curie constant~\cite{Zhang2017arxiv}. 
This fit is better expressed by the linear behavior of the inverse
susceptibility, $\chi^{-1}_{B_{2g}}(0,T,x)$. 
The values of
$T_\theta(x)$, which correspond to the abscissae of the linear fits,
are given in Table \ref{TS} and shown in Fig.~\ref{Fig4_summary}
\footnote{We note that the derived $T_\theta(x)$ and $Q(x)$ depend
on the background subtraction. 
Here we used the $X'Y'$ background
as a reference. The error bars for $T_\theta(x)$ are estimated from
the fitting procedure.}. 

$T_\theta(x)$ is monotonically decreasing function of  
$x$, at about 40-60~K lower than $T_S(x)$. 
At a critical doping $x_{c}=0.031$ the Weiss temperature approaches zero, 
indicating a hidden quantum critical point. 
At this critical Au-doping concentration the system is a 
superconductor with $T_{c}=2.5$~K~\cite{Li2015PRB}, indicating a 
strong connection  
between quantum critical nematic fluctuations and unconventional 
superconductivity~\cite{Chu824science2010,Lederer2015PRL}. 

For pristine BaFe$_2$As$_2$, susceptibility 
$\chi_{B_{2g}}(0,T,x=0)$ is rapidly suppressed just below $T_{S}/T_{N}$. 
In contrast, for the Au-doped samples,  $\chi_{B_{2g}}(0,T,x>0)$
has a plateau-like saturation between split 
$T_{S}(x)$ and $T_{N}(x)$, with only a slight decrease, followed by 
much faster decrease below $T_{N}(x)$ to a saturation value at 
temperature coinciding with magnetic order parameter saturation 
temperature, 
Figs.~\ref{Fig3_Chi0}(e)-\ref{Fig3_Chi0}(h).

The fact that the static susceptibility stops increasing on cooling below $T_{S}$, 
when $T_{N}$ is lower than $T_{S}$, indicates that the origin of
the critical fluctuations is not magnetic but rather is driven 
by the charge quadrupole
fluctuations~\cite{Thorsmolle2016PRB,ZhangWL2014arxiv,Zhang2017arxiv}. 
The
likely scenario for quadrupole charge fluctuations with $B_{2g}$
symmetry is a charge transfer between the nearly degenerate $d_{xz}$
and $d_{yz}$ orbitals. 
Such quadrupole charge fluctuations are expected to slow down below 
$T_{S}$, when the degeneracy of $d_{xz}$/$d_{yz}$ orbitals is lifted, 
and be suppressed below $T_{N}$, when the SDW gap depletes 
the electronic DOS at the Fermi level, with complete suppression when
the magnetic ordering is fully established.

In more details, the electron transfer between quasi-degenerate
$d_{xz}/d_{yz}$ orbitals induces a charge quadrupole moment $Q(x)$
proportional to the local charge imbalance $n_{xz}-n_{yz}$. 
The charge quadrupole moment can be estimated from the fitted Curie
constant, Eq.~(\ref{Curie}). 
The $Q(x)$ values for
Ba(Fe$_{1-x}$Au$_x$)$_2$As$_2$ are summarized in Table~\ref{TS}.  
In the inset of Fig.~\ref{Fig4_summary}, we show the doping dependence of $Q(x)$
and of the ordered magnetic moment $M(x)$ as a function of Au
concentration $x$. 
The opposite doping dependence of the charge quadrupole
moment $Q(x)$ and of the magnetic moment $M(x)$ suggests that the
charge quadrupole order at the Fe site competes with stripe magnetic
order. Same conclusion can be reached from the fact that the static
$XY$ quadrupole susceptibility saturates at $T_S$ and is gradually
suppressed below $T_N$ as the stripe magnetic order parameter builds
up.

In conclusion, we studied nematic Raman response for Au-doped 
BaFe$_{2}$As$_2$ samples, which 
have split structural and magnetic transition temperatures. 
The data revealed that above $T_{S}$ 
the static Raman susceptibility in the quadrupole 
$B_{2g}$ channel 
follows a Curie-Weiss behavior.
Between $T_{S}$ and $T_{N}$, the susceptibility stops increasing, and
it decreases only below $T_N$, when the magnetic order
parameter develops and a SDW gap opens. 
We attribute the
corresponding increase of the Raman susceptibility to quadrupole
charge fluctuations due to electron transfer between nearly
degenerate $d_{xz}$ and $d_{yz}$ orbitals. 
Weiss temperature $T_\theta(x)$ extrapolated from the
high-temperature fluctuations is decreasing with doping concentration 
$x$ and is several tens of degrees lower than $T_S(x)$ and $T_N(x)$. 
Importantly, the critical concentration $x_{c}$ defining the quantum 
critical point $T_\theta(x_{c})=0$ is located inside the small 
superconducting dome, thus underlining the 
role played by the quadrupole orbital fluctuations
in the pairing mechanism of the Fe-based
superconductors~\cite{Chu824science2010,Lederer2015PRL}. 
The extrapolated charge quadrupole moment $Q(x)$ is monotonically 
increasing function, while the ordered magnetic moment $M(x)$ is a 
decreasing function of $x$, suggesting that the charge quadrupole
order competes with the collinear antiferromagnetic order.

The spectroscopic research at Rutgers was supported by the US Department of Energy,
Basic Energy Sciences, and Division of Materials Sciences and
Engineering under Grant No. DE-SC0005463. 
The sample growth and characterization at ORNL was
supported by the US Department of Energy, Basic Energy Sciences,
Materials Sciences and Engineering Division. Work at IOP was
supported  by grants  from NSFC (11674371 and 11274362) and  MOST
(2015CB921301, 2016YFA0401000 and 2016YFA0300300) of China.

\end{document}